\documentclass[12pt]{article}
\usepackage{latexsym}
\newlength{\singlespace}
\newlength{\doublespace}
\newcommand{\beq}{\begin{equation}}
\newcommand{\eeq}{\end{equation}}
\def\la{\mathrel{\mathchoice {\vcenter{\offinterlineskip\halign{\hfil
$\displaystyle##$\hfil\cr<\cr\sim\cr}}}
{\vcenter{\offinterlineskip\halign{\hfil$\textstyle##$\hfil\cr
<\cr\sim\cr}}}
{\vcenter{\offinterlineskip\halign{\hfil$\scriptstyle##$\hfil\cr
<\cr\sim\cr}}}
{\vcenter{\offinterlineskip\halign{\hfil$\scriptscriptstyle##$\hfil\cr
<\cr\sim\cr}}}}}

\makeatletter

\def\thebibliography{
\section*{References}
\def\leftmark{Bibliographie} \def\rightmark{Bibliographie}
\list{}{\labelwidth 0pt \leftmargin 0em \labelsep 0pt
 \parsep 0pt \itemsep 1em plus 1pt minus 1pt \listparindent 0pt
 \itemindent -1.5em\listparindent \itemindent
 \rightmargin\leftmargin\advance\leftmargin 1.5em}
\def\csoldstn##1{##1}
\def\csoldstnb##1{\bf##1}
\def\rng##1-##2{{\csoldstn{##1}}\lower0.2ex\hbox{--}{\csoldstn{##2}}}
\def\oo##1{{\csoldstn{##1}}}
\def\and{and }

\def\pgs{pp.}
\def\pag{p.}
\def\in ##1/{In {\it ##1,}}
\def\ed ##1/{(##1, Ed.)}
\def\eds ##1/{(##1, Eds.)}
\def\su ##1/{submitted to ##1,}
\def\ip ##1/{##1, in press,}
\def\up ##1/{##1, unpublished.}
\def\iprep /{in preparation}

\def\co ##1/{}
\def\a ##1/{\item{\sc ##1}}
\def\t ##1//{##1.}
\def\tl ##1##2/{{\sl \uppercase{##1}\lowercase{##2},}}
\def\bt ##1/{in\ {\it ##1,}}
\def\st ##1/{##1,}
\def\pu ##1/{##1,}
\def\j ##1/{{\it ##1}}
\def\vo ##1/{{\csoldstnb##1,}}
\def\yp ##1 ##2-##3/{({\csoldstn{##1}}), \pgs~\rng{##2}-{##3}.}
\def\py ##1-##2 (##3)/{({\csoldstn{##3}}), \pgs~\rng{##1}-{##2}.}
\def\ysp ##1 ##2/{({\csoldstn{##1}}), \pag~{\csoldstn{##2}}.}
\def\psy ##1 (##2)/{({\csoldstn{##2}}), \pag~{\csoldstn{##1}}.}
\def\p ##1-##2/{\rng{##1}-{##2}.}
\def\bp ##1-##2/{pp.\ \rng{##1}-{##2}.}
\def\ps ##1/{{\csoldstn{##1}}.}
\def\y ##1/{{\csoldstn{##1}}.}
\def\nu ##1/{{\csoldstn{##1}},}

\sloppy \interlinepenalty\@M
\sfcode`,=1000 \sfcode`:=1000 \sfcode`;=1000
\sfcode`.=1000 \sfcode`?=1000 \sfcode`!=1000
}

\makeatother

\def \al{{{\it et al.}}}
\def \gcm3{{\rm \> g}^{\ }{\rm cm}^{-3}}
\def \e16{\times 10^{16} {\rm \> kg}}

\def \approxless{\ \lower.5ex\hbox{$\sim$}\raise.5ex\llap{$<$}\ }
\def \approxgreater{\ \lower.5ex\hbox{$\sim$}\raise.5ex\llap{$>$}\ }

\title{KBO binaries: are they really primordial ?}
\author{
 \\
\\
J-M. Petit$^1$, O. Mousis$^1$
\\[3mm]
\\
$^1$Observatoire de Besan\c{c}on, France\\
\\
\\
\\
Proposed running head:  \\
\\
\\
Submitted to Icarus, Mai 9th, 2003 
\\
\\
\\
14 text pages + 3 figures + 3 tables = 20 pages in initial submission
\\
}
\date{\rule{0mm}{0mm}}

\begin{document}
\maketitle

\setlength{\baselineskip}{\singlespace}

\newpage

\vspace*{2.5cm}

\begin{center}
{\bf Abstract}
\end{center}

\vspace*{2.5cm}

Given the large orbital separation and high satellite-to-primary mass
ratio of all known Kuiper Belt Object (KBO) binaries, it is important to
reassess their stability as bound pairs with respect to several
disruptive mechanisms.
Besides the classical shattering and dispersing of the secondary due to a
high-velocity impact, we considered the possibility that the secondary is
kicked off its orbit by a direct collision of a small impactor, or that it
is gravitationally perturbed due to the close approach of a somewhat larger
TNO.

Depending on the values for the size/mass/separation of the binaries that
we used, 2 or 3 of the 8 pairs can be dispersed in a timescale
shorter than the age of the solar system in the current rarefied
environment.
A contemporary formation scenario could explain why we still observe these
binaries, but no convincing mechanism has been proposed to date.
The primordial formation scenarios, which seem to be the only viable ones,
must be revised to increase the formation efficiency in order to account
for this high dispersal rate.
Objects like the large-separatioKBO binary n2001~QW$_{322}$ must have been
initially an order of magnitude more numerous.

If the KBO binaries are indeed primordial, then we show that the
mass depletion of the Kuiper belt cannot result from collisional grinding,
but must rather be due to dynamical ejection.

\vspace{0.5cm}


\newpage
\section{Introduction}

Over the past decade, the Edgeworth-Kuiper belt has changed status, from a
theoretically predicted entity to a collection of more than 700 comets
orbiting beyond Neptune.
At first, those (not so) small icy bodies were thought to be lonely
wanderers, except for the pair Pluto-Charon.
At the end of 2000, Veillet \al\ (2002) found the first Kuiper Belt Object
(KBO) satellite.
This discovery was followed by seven others in the following 24 months,
representing about 1\% of the total known KBO population.
The main characteristics of the KBO binaries, when compared with the
asteroid binaries, are large separations ($\sim$5,000 to 130,000~km, or
$\sim$20 to almost 2,000 times the primary radius - of order a few to 10
for asteroids) and high satellite-to-primary mass ratio of 0.1 to 1
($\sim$10$^{-4}$ to 10$^{-3}$ for asteroids).

The set of known KBO binaries suffers from a very strong observational
bias.
KBO binaries with a small separation are impossible, or at least very
difficult to detect as binaries because of their large distance to Earth.
Their angular separation is smaller than the typical seeing, and still
smaller than the diffraction limit (achievable with Adaptive Optics) if the
separation is comparable to that of the asteroid binaries.
Likewise, KBO binaries with low satellite-to-primary mass ratios cannot
be recognized as binaries, because the secondary falls beyond the
limiting magnitude of most observations.
However, the very existence of the known binaries is a great novelty with
respect to what is known in the asteroid belt or in the NEO population.
This has prompted several authors to study their formation mechanisms
(Goldreich \al, 2002; Stern, 2002; Weidenschilling, 2002).
Goldreich \al\ and Weidenschilling concluded that collisions in the
current Edgeworth-Kuiper cannot account for the large number of binaries
found, nor for their large separation and high satellite-to-primary mass
ratios.
They proposed various mechanisms that must have occured in the late stage of
the formation of the solar system, at the end of the accretion phase.
According to Goldreich \al\ and Weidenschilling, the binaries would be
primordial.
Although contemplating similar primordial scenarios, Stern favors
more contemporary collisional formation mechanisms, and reconcile the
number of required impactors with the actual number of bodies by assuming a
surface albedo of the binaries to be $\sim$15\%, 2 to 4 times larger than
usually assumed.

Once formed, a binary object can disappear either because one of the
components (usually the secondary) is destroyed (shattered and dispersed)
through a high velocity impact, or the pair gains enough orbital energy to
become unbound, due to the close approach or direct collision of another
object.
For asteroids, the major mechanism to eliminate a binary is the destruction
of the secondary through high-velocity impacts.
Since it seems well established that all known KBO binaries cannot be
efficiently collisionally destroyed in less than 4~Gyr, all work to date
have assumed that the KBO binaries would be stable over the age of the
solar system, except for Weidenschilling (2002), who mentioned, without any
development, the possibility of disrupting the most loosely bound binaries.
We show that long term stability is not guaranteed, and some of the KBO
binaries may very well have lifetimes of order 1-2~Gyr.

In the present work, we estimate the stability of these binaries with
respect to several dispersal mechanisms.
The data describing the known binaries and their dynamical and
collisional environments are listed in Sect.~\ref{s:facts}.
Besides the classical shaterring and dispersing of the secondary through a
direct collision, we also consider the possibility that the secondary is
knocked off its orbit by a direct collision of a rather small impactor, or
is gravitationally perturbed by the close approach of a somewhat larger
TNO.
All these mechanisms are described in Sect.~\ref{s:mechanisms}.
In Sect.~\ref{s:lifetimes}, we present the lifetimes of the KBO
binaries with respect to all three disruption mechanisms, both in
the current rarefied and in the denser initial environments.
We discuss the implications of these results on the formation
scenarios in Sect.~\ref{s:discussion}.
Finally, a summary of our findings is given in Sect.~\ref{s:summary}.

\section{The facts}
\label{s:facts}

To address our goal, we first need to know the parameters defining the
binaries, and then the population of potential impactors.
The binary parameters we use here (Table~\ref{t:characteristics}) are from
two different compilations for the first seven of them, the first one by
Merline \al\ (2003), the second one by Stern (2002), yielding different
sizes, masses, and separations.

The last binary 2001~QC$_{298}$, not included in these compilations, was
discovered in October, 2002, and reported by Noll \al\ (2003).
Very little information is given in the discovery announcement.
From the published magnitude and distance, we estimated the equivalent
radius of the pair ${\rm R}_{eq} = \sqrt{{\rm R}_P^2 + {\rm R}_S^2}$ to be
212~km, assuming an albedo of 0.04, the usual default value for KBOs.
The separation projected on the sky is estimated to be 5000~km $\pm$
2000~km.
The difference in magnitude is not known to us.
Hence we chose two different cases, at the limits of the interval for known
binaries: zero magnitude difference at one end of the range, and 2.2
magnitude difference at the other extreme, the largest known magnitude
difference for KBO binaries.
The resulting parameters are displayed in last line of
Table~\ref{t:characteristics}.

The number of objects in the Edgeworth-Kuiper belt is not yet very well
known.
For the sake of simplicity and to allow comparison with previous work, we
use the same differential size distribution as proposed by Weissman and
Levison (1997), and Durda and Stern (2000), {\it i.e.} a two-component
power law of the form:
\begin{equation}
N(r_i) \propto r_i^b {\rm d}r_i
\label{e:distrib}
\end{equation}
where $b = -3$ for $r < r_0$ and $b = -4.5$ for $r >
r_0$, with $r_0 = 5$~km.
The differential size distribution is assumed to be continuous at $r =
r_0$.
Following Durda and Stern (2000), the normalization constant should be at
least 70,000 objects with radius larger than 50~km, and perhaps twice that
many.
So we use $10^5$ objects larger than 50~km in radius.

The final piece we need to estimate the number of collisions on a given
target from a given set of impactors is the intrinsic collision probability.
This number depends on the actual orbital distribution of the TNOs, and is
therefore not well determined.
It also depends on the location of the target in the belt.
Here, we use the average value proposed by Farinella \al\ (2000)
\begin{equation}
\langle P_i \rangle = 1.3 \, \times \, 10^{-21} \> {\rm km}^{-2}\, {\rm
yr}^{-1}.
\label{e:prob-col}
\end{equation}

\section{Disruption mechanisms}
\label{s:mechanisms}

\begin{figure}[htbp]
\hspace{25mm} a \hspace{95mm} b

\vspace{50mm}
\includegraphics{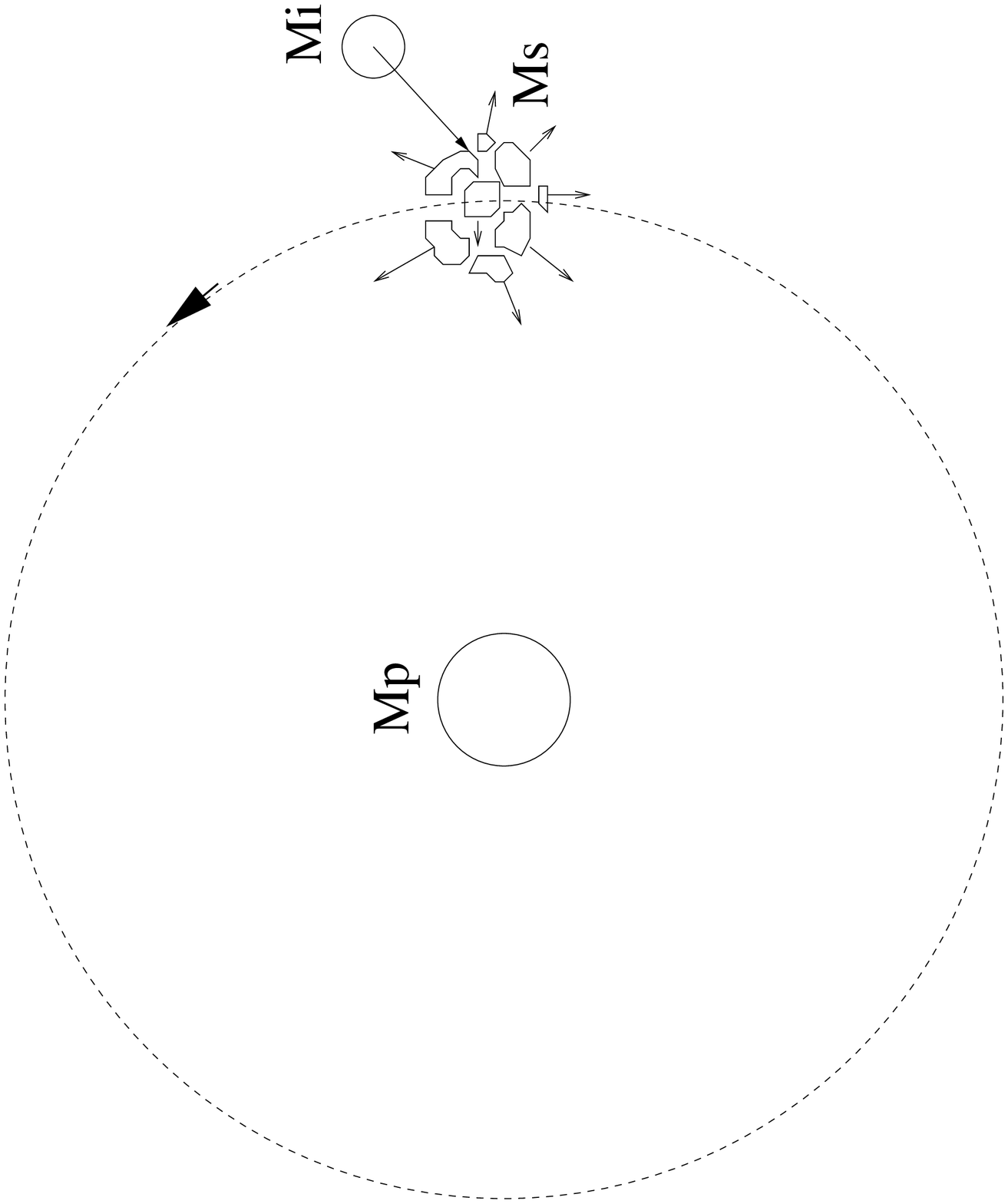}
\includegraphics{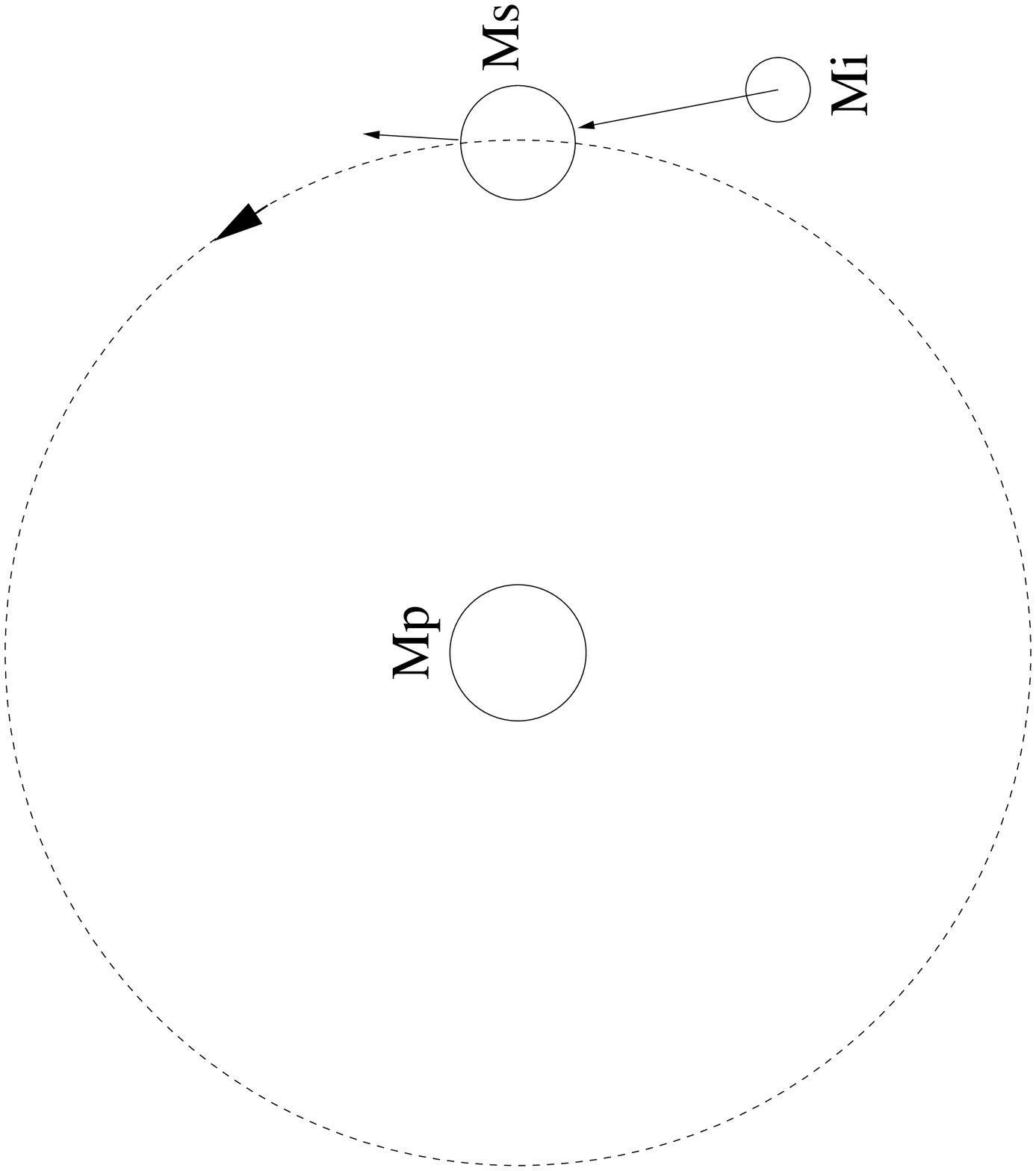}
\hspace{75mm} c

\vspace{50mm}
\includegraphics{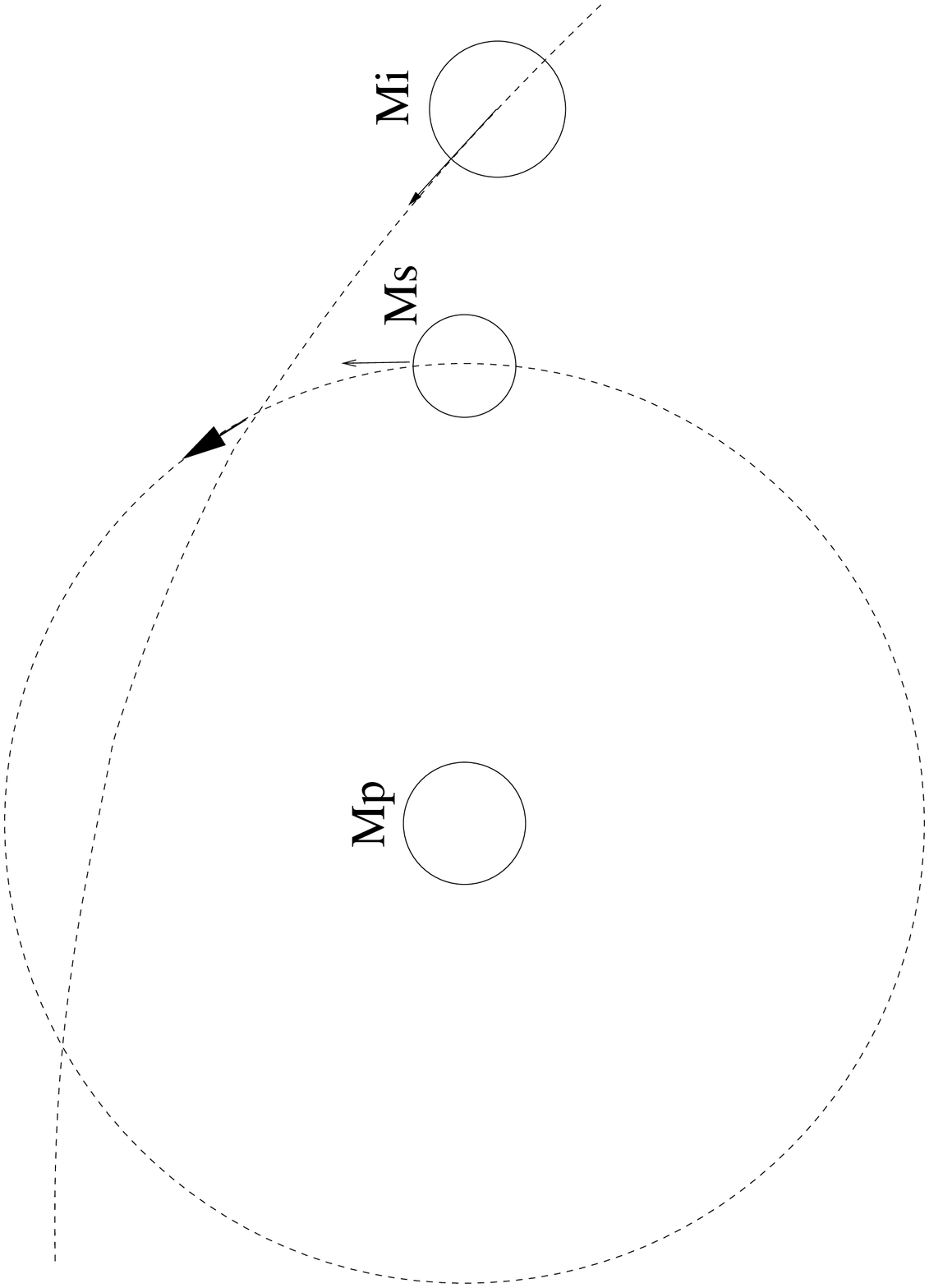}
\caption{
(a): The secondary is shattered and its fragments are dispersed due
to a high-velocity impact.
(b): The secondary is kicked off its orbit around the primary due
to a direct collision by another TNO.
(c): The secondary is dislodged from its orbit around the primary
due to the gravitational perturbation from a passing TNO.
}
\label{f:disrupt}
\end{figure}

\begin{figure}[htbp]
\vspace{70mm}
\includegraphics{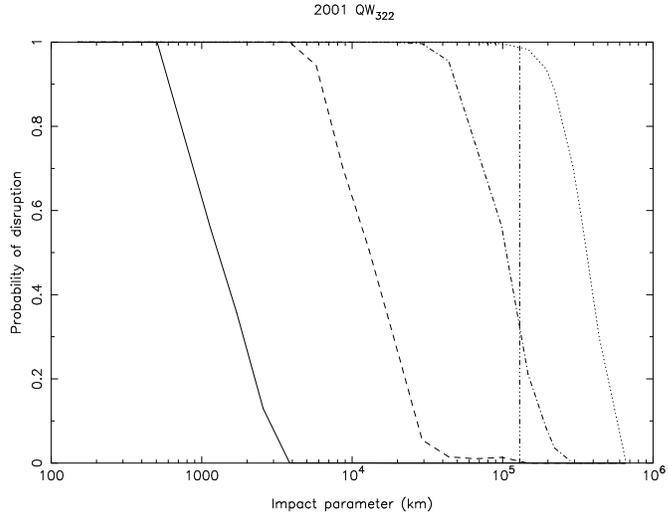}
\caption{
Probability of disruption of 2001~QW$_{322}$ due to a projectile arriving
at 500~m/s with mass of 10$^{19}$ (solid line), 10$^{20}$ (dashed line),
10$^{21}$ (dash-dotted line) and  10$^{22}$~kg (dotted line).
The dash-triple dot line indicates the orbital separation of
2001~QW$_{322}$.
}
\label{f:grav-qw322}
\end{figure}

In this work, we consider three different ways  (Fig.~\ref{f:disrupt}) of
eliminating a KBO binary.
\begin{itemize}
\item The first one is the shattering of the secondary by a
collision, followed by the dispersing of the resultant fragments.
This possibility has been studied at length in previous works, in
particular in the framework of the asteroid belt.
Davis and Farinella (1997) show that bodies of radius larger than
50~km cannot be shattered and dispersed in the current dynamical and
collisional environment.
Since all binaries considered here have a secondary larger than 50~km in
radius, it is clear that this process cannot be an efficient mechanism
for eliminating the known KBO binaries.
However, we consider this case as a reference, and as a mean of comparison
with the other mechanisms.
We use the value of $Q^*_D$  for ice (minimal energy per unit mass
of target to shatter and disperse the target) given by Benz and Asphaug
(1999) to compute the required minimum impactor size:
\begin{equation}
Q^*_D = Q_0 \left(\frac{R_{pb}}{1cm}\right)^{\alpha}
           + B \rho \left(\frac{R_{pb}}{1cm}\right)^{\beta}.
\label{e:q-star-d}
\end{equation}
$R_{pb}$ is the radius of the {\it parent body} to shatter and disperse,
expressed in cm and $\rho$ is the density of the parent body (in g/cm$^3$).
$\alpha$, $\beta$, $B$ and $Q_0$ are constants determined by a fit of
results of numerical experiments, for impact velocity of 500~m/s and
3000~m/s.
Since we will use impact velocities of 500~m/s and 1500~m/s (see below),
the values of the parameters for the latter case are derived by linear
interpolation from those given by Benz and Asphaug.

\item The second mechanism is the collision of a small projectile, not big
enough to shatter the secondary, but that gives enough momentum to unbind
the secondary from the primary (Fig.~\ref{f:disrupt}b).
For all known KBO binaries, it is easier to unbind the secondary than to
send it colliding with the primary, i.e. $e \> \to \> 1$.
When an impactor of mass $M_i$ hits the secondary of mass $M_S \> >> \>
M_i$, the secondary undergoes a change in velocity of ${\bf \Delta V} = M_i
{\bf V_i} / M_S$ where ${\bf V_i}$ is the impactor's relative velocity.
At this point, it is convenient to introduce the total mas of the binary,
$M = M_P + M_S$, where $M_P$ is the mass of the primary, and the reduced
mass $\mu = M_P M_S / (M_P + M_S)$.
Before the kick, we assume the secondary to be on a circular orbit around
the primary, with speed $V_S = \sqrt{G M/r}$, where $G$ is the
gravitational constant, and $r$ the separation between the primary and the
secondary.
The velocity after the kick is ${\bf V'_S} = {\bf V_S} + {\bf \Delta V}$.
We look for a value of that velocity such that the total energy of the
system vanishes, that is:
\begin{equation}
\frac{1}{2} \mu (V'_S)^2 = \frac{G M_P M_S}{r} = \mu (V_S)^2
\end{equation}
(circular inital orbit).
The square modulus of the velocity is given by
\begin{equation}
(V'_S)^2 = (V_S)^2 + (\Delta V)^2 + {\bf V_S} \cdot {\bf \Delta V}
         = (V_S)^2 + (\Delta V)^2 + V_S \Delta V \cos{\theta},
\end{equation}
where $\theta$ is the angle between the impactor's and the secondary's
velocities.
Averaging over all impact directions, we obtain
\begin{equation}
\langle \Delta V \rangle = V_S \frac{\sqrt{5} - 1}{2} \simeq 0.62
\sqrt{\frac{G M}{r}}.
\label{e:dv}
\end{equation}
So finally, the average impactor's mass necessary to dislodge the secondary
from its orbit by direct collision is
\begin{equation}
M_i = 0.62 \frac{M_S}{V_i} \sqrt{\frac{G M}{r}}.
\end{equation}

\item The last possibility is gravitational perturbation from an encounter
with a third body, that will transfer enough energy to the binary to
unbind it (Fig.~\ref{f:disrupt}c).
We have performed numerical integrations of the 3-body problem to determine
the unbinding gravitational cross section for a perturber of mass $M_i$ =
10$^{19}$, 10$^{20}$, 10$^{21}$ and  10$^{22}$~kg, with velocity $V_{i}$.
For each value of the mass and incoming velocity, we have selected a set of
impact parameters, from 150 to 660,000~km, with 1.5 ratio increments.
For each impact parameter, we ran 10,000 simulations with all other
parameters taken at random, to evenly sample the space of possible
orientation.
Integrations were performed using the well-known general purpose,
self-adaptative Bulirsh--Stoer integrator (Stoer and Bulirsch, 1980) with
relative precision of 10$^{-12}$.
From this we determined the probability of disruption of the binary as a
function of the impact parameter.
Fig.~\ref{f:grav-qw322} shows this probability for 4 different masses of
the projectile (10$^{19}$, 10$^{20}$, 10$^{21}$ and  10$^{22}$~kg) arriving
at 500~m/s on 2001~QW$_{322}$.
This case has been chosen as being representative of all cases, with no
particular behavior.
The probability of disruption $P(h)$ for impact parameter $h$ determines
the gravitational disruption cross-section
\begin{equation}
\sigma = \int_0^\infty 2 \pi h P(h) {\rm d}h,
\label{e:grav-cross-section}
\end{equation}
from which one can derive the frequency of occurence of such disruptions,
and finally define the equivalent radius R$_g$ = $\sqrt{\sigma/\pi}$.
Note that on Fig.~\ref{f:grav-qw322}, the distance between curve decreases
between the last two on the right.
This results in a maximum efficiency (minimum lifetime) for the
gravitational disruption mechanism somewhere in the range of mass studied.

\end{itemize}

\section{Lifetimes}
\label{s:lifetimes}

In order to determine the frequency of disruption events, or conversely the
expected lifetime with respect to disruption, one need to know the number of
projectiles, the disruption cross-section, and the intrinsic encounter
probability.
Given the size of the projectile, one can easily determine the number of
such projectiles using the size distribution given by
eq.~(\ref{e:distrib}).
For the first two disruption mechanisms (direct collision), the disruption
cross-section is simply the collisional cross-section, that is the
physical cross-section $\pi ({\rm R}_S + {\rm R}_i)^2$ times the
gravitational focussing $(1 + V_{esc}^2/V_{i}^2)$, where $V_{esc}$ is
the escape velocity of the pair (secondary, impactor).
However, the $\pi$ factor is already included in the definition of $\langle
P_i \rangle$.
Hence we only need to compute $({\rm R}_S + {\rm R}_i)^2$ times the
gravitational focussing.
For the third disruption mechanism, we compute $\sigma/\pi$ from
eq.~(\ref{e:grav-cross-section}).

For each set of binary parameters (Merline \al, 2003 and Stern, 2002), we
have estimated the impactor size and/or the disruption cross-section for
the three mechanisms, assuming encounter velocities of 500~m/s and
1500~m/s which roughly bracket the actual encounter velocities in the
present day Edgeworth-Kuiper Belt.
In Table~\ref{t:lifetimes} we report the shortest lifetime and the
corresponding impactor size for each mechanism for the seven KBO binaries
listed by Merline \al\ (2003) and Stern (2002).

The case of 2001~QC$_{298}$ is presented in Table~\ref{t:lifetimes-qc298}.
Here we have considered four different sets of binary parameters: the two
sets of radii from the last line of Table~\ref{t:characteristics}, and a
density of either 1~g/cm$^3$ or 2~g/cm$^3$.
For each of these sets, we have run simulations for the same encounter
speeds as before, and we report the shortest lifetime for each parameter
set.

The first obvious trend is that ejection of the secondary due to a
direct collision is the most efficient way to disrupt a KBO binary.
As was already well known, we find that collisional shattering and
dispersing of the secondary is not efficient here.
Gravitational disruption is also inoperative here because of the large size
needed for the projectile, and the steep slope of the size distribution at
large sizes.
Interestingly, thanks to the high encounter speed, we never saw an exchange
between the projectile and one of the components of the binaries in any of
our integrations.

As can be seen from the tables, 2 or 3 of the 8 known KBO binaries
have mean disruption lifetimes shorter than the age of the solar system,
even if all secondaries would survive shattering disruption over that time
span.
2001~QW$_{322}$ cannot survive in its current state for more than
1 to 2~Gyr.
2000~CF$_{105}$ would most certainly have been destroyed if it
was primordial.
The case of 1998~WW$_{31}$ is not completely settled yet.
Depending on the exact parameters for its components and the relative
orbit, it may or may not survive for 4~Gyr.

Up to now, we have solved the disruption equation for a single encounter.
Since the number of small impactors is larger than the number of large
impactors, we must also consider the effect of multiple collisions by small
impactors on the secondary.
In this case, the secondary will experience a random walk.
The total change in velocity will grow like
\begin{equation}
\Delta V  = \sqrt{\sum (\delta V)^2},
\label{e:dvmulti}
\end{equation}
where $\delta V$ is the change of velocity due to each collision from a
small impactor of mass $m_i$.
As before, $\delta V \propto m_i$, and the number of collisions, in a fix
timespan, is proportional to the number of impactors of mass $m_i$,
$n(m_i)$.
In the following, we only consider a single power law size distribution,
meaning that we will only be able to compare lifetimes or efficiency for
masses on the same side of $r_0$.
From the differential size distribution of eq.~(\ref{e:distrib}), the
differential mass distribution is:
\begin{equation}
n(m_i) \propto m_i^{\frac{b-2}{3}} {\rm d} m_i.
\label{e:mdistrib}
\end{equation}
So the effect of collisions from impactors of mass $m_i$ varies like
\begin{equation}
\Delta V \propto m_i^{\frac{b+4}{6}}.
\label{e:mdv}
\end{equation}
Hence for $b > -4$, the largest impactors have the dominant effect, while
for $b < -4$, the cumulative effect of small impactors overcomes the
effect of a single collision by a big impactor.
It is important to note that the effect on the velocity of the secondary is
a continuous function of the impactor's mass, while the collisional erosion
rate exhibits a large discontinuity for masses smaller than the critical
mass for shattering and dispersing the secondary.
The size distribution we have used so-far has $b = -4.5$ in the range of
sizes of the disruptive impactors (Tables~\ref{t:lifetimes} and
\ref{t:lifetimes-qc298}).
Hence impactors of size $r_0$ would be collectively more efficient at
disrupting the binaries.
Since for $r_i < r_0$, $b = -3$, smaller projectiles would be less
efficient at disrupting the binaries.
Egaling eq.~(\ref{e:dv}) and (\ref{e:mdv}), and noting that the change in
velocity is proportional to the square root of the timespan, we relate the
disruption lifetime $T_s$ due to multiple collisions from bodies of size
$r_0$ to the one ($T_l$) computed earlier for a single collision:
\begin{equation}
T_s = T_l \left( \frac{m_l}{m_s} \right)^{\frac{b+4}{3}}
\label{e:lifetimesmall}
\end{equation}
where $m_l$ is the mass of the large impactor, and $m_s$ the mass of the
small impactors.
Here, we have used the fact that the impactors are always small compared to
the secondary, and then only the mass of the secondary governs the
gravitational focusing.
This reduces the lifetimes given in Tables~\ref{t:lifetimes} and
\ref{t:lifetimes-qc298}, although not in a way that changes our previous
conclusions.
The same 3 binaries are disrupted, maybe a little faster, and the other one
can still survive for the age of the Solar System.

A word of caution is in order here.
Dynamical friction from a swarm of small bodies has been said to cause a
hardening of the binaries.
It is not clear that bodies of radius $r_0 = 5$~km actually participate in
the dynamical fristion, hence hardening the binaries instead of disrupting
them.
But the single disruptinve collisions still occur on the time scales given
in Tables~\ref{t:lifetimes} and \ref{t:lifetimes-qc298}, which then set an
upper limit for the lifetimes.

In the previous calculations, we have considered a population of
projectiles corresponding to today's Edgeworth-Kuiper belt.
However, it seems most likely that the primordial belt had to be much more
massive, as much as 100 times more massive, in order to grow bodies as
large as those observed today (Stern, 1996).
The increase in mass can be achieved by simply multiplying the number of
objects of each size by a constant factor of order 100, retaining the same
size distribution, or by keeping the same number of large bodies, and
increasing the mass in small bodies.
Some authors (Stern and Colwell, 1997; Davis and Farinella, 1997) have
argued that the mass loss of the Kuiper belt is due to collisional erosion.
From our previous estimates, we can see that a long lasting intense
collisional activity can have profound effects on the KBO binaries.
We now investigate these effects on the direct collision ejection
mechanism.

\begin{figure}[htbp]
\vspace{70mm}
\includegraphics{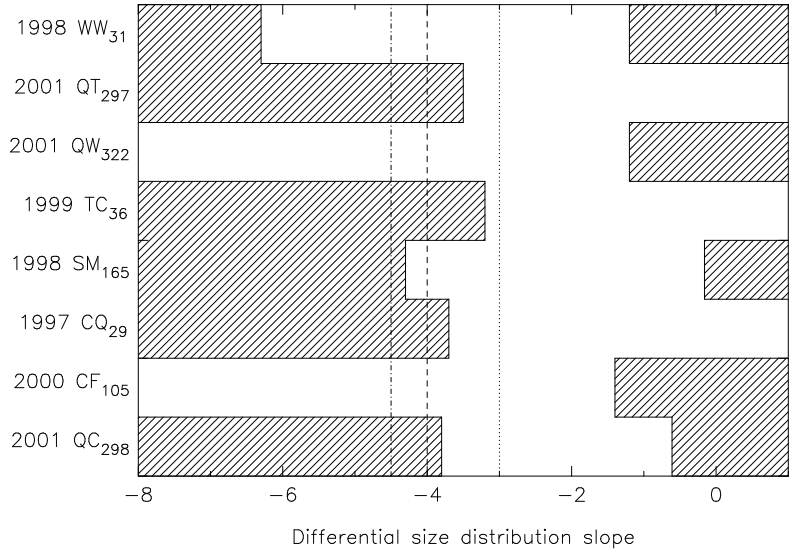}
\caption{
The hashed regions correspond to slopes of the differential size
distribution for which collisional disruption of a binary occurs less
frequently than shattering and dispersing of a 50~km body in the massive
primordial environment.
The -4 slope (dashed line) is the limit below which multiple collisions of
small impactors are more efficient than single collisions of larger
impactors.
The dash-dotted line corresponds to the large-end size distribution
exponent in eq.(\ref{e:distrib}), the dotted line corresponding to the
small-end.
}
\label{f:slopes}
\end{figure}

We now suppose that the mass loss of the belt is actually due to
collisional grinding.
In this case, both Davis and Farinella (1997) and Stern and Colwell
(1997) concluded that all primordial bodies of radius 50~km or less
would have been destroyed, the ones we see now being fragments due
to the shattering of bigger parent bodies.
For each density and impact velocity assumed so-far, we can estimate
the minimum size of an impactor capable of shattering and dispersing a
50~km radius body from eq.~(\ref{e:q-star-d}).
We compare this size and the corresponding collisional cross-section of a
body of 50~km radius to the size and collisional cross-section of an
impactor large enough to push the secondary out of its orbit, as in our
second disruption mechanism.
The occurence frequency of these two types of events is the product
of the collisional cross-section time the number of impactors time
the intrinsic collision probability.
The intrinsic collision probability has changed over the age of the
solar system, and cannot be given by eq.~(\ref{e:prob-col}) at all
times, but it is the same for both types of events at any times.
So we do not need to know its value to compare the frequencies.
We just need to compare the cross-sections and numbers of impactors.
Assuming a power-law size distribution like in
eq.~(\ref{e:distrib}), we can derive a condition on the slope $b$ so
that a KBO binary would be disrupted by a direct impact more often
than a 50~km radius body would be shattered and dispersed.
This corresponds to the open region in Fig.~\ref{f:slopes}, while the
hashed regions correspond to slopes for which a KBO binary would be
disrupted by direct impact less frequently.
The current slope for large bodies, -4.4$\pm$0.3 (Gladman \al, 2001), is a
relic of the accretion phase.
Later collisional evolution tend to push the slope toward -3.5 or even -3,
starting with the small bodies.
So clearly KBO binaries like 1998~WW$_{31}$, 2000~CF$_{105}$ and
2001~QW$_{322}$ (large orbital separation) cannot survive an intense
initial collisional activity.
Even a large fraction of objects like 1998~SM$_{165}$ would be disrupted.

The four remaining binaries could, in some cases, resist disruption even
with a size distribution shallower than $b = -4$.
This would happen only if all collisions occur at high speed
(1500~m/s).
Large speeds favor shattering and dispersing over ejection since the
former depends on the square of the velocity, while the latter
depends on the velocity.
1997~CQ$_{29}$, 1999~TC$_{36}$ and 2001~QC$_{298}$ are all very
close binaries, with a large secondary, increasing their stability.
2001~QT$_{297}$ is a rather well separated binary (20,000~km), but
has the largest of all secondaries.

\section{Discussion}
\label{s:discussion}

The existence of an object like 2001~QW$_{322}$, with a lifetime of
1 to 2~Gy in the current rarefied environment means than there were,
at least, between 7 and 50 similar KBO binaries 4~Gyr ago.

As expected, the most largely separated are the easiest to disrupt.
In other words, the KBO binaries easiest to disrupt are also the ones that
are the most difficult to create in Goldreich \al\ (2002; L$^2$s for two
large bodies and a sea of small bodies) and Stern (2002; lcL$^2$ for late
collision of two large bodies) scenarios.
Weidenschilling (2002; cL$^2$L for collision of two large bodies in the
vicinity of a third large body) scenario, on the other hand, tends to
create more large separation binaries.
Since large separation binaries are more prone to disruption, this could
reconcile cL$^2$L with the observations, showing more binaries with small
separations (4 with a distance 5000-8000~km, 3 with a distance
20,000-23,000~km and only 1 with a distance $>$~100,000~km) than with
large separations.

L$^2$s tend to form enough large separation binaries, if considering only
their current number.
However, since one must assume that there were at least an order of
magnitude more large separation binaries 4~Gyr ago, this implies that at
least the same increase in formation frequencies occured for small
separation binaries.
This means that binaries like 1998~SM$_{165}$ or 1999~TC$_{36}$ should be
at least 10 times more numerous.
One possible explanation as to why this is not reflected in the current
sample is that such binaries already suffer from a strong observational
bias.
The HST survey for KBO binaries will bring important information to try
and answer this question.

An other possibility is that both L$^2$s and  cL$^2$L have been active to
form KBO binaries, L$^2$s forming all the small separation ones, while
cL$^2$L produced a large number of large separation binaries.
Finally, the short lifetime of the latter would explain the current
distribution of KBO binaries.

The Stern (2002) scenario is more difficult to evaluate, partly because
the main formula (left column of p. 2301 in that paper) is obviously
dimensionaly wrong.
Hence it is difficult to weigh the importance of each assumption.
However, Stern favors a recent formation of the KBO binaries, arguing that
the binary albedo could be as much as 4 times larger than usually
assumed.
In this case the size of the binary components would be divided by 2, and
therefore the required impactor mass would decrease by almost an order of
magnitude.
This in turn would increase drastically their number.
This last part of reasoning is true only if the albedo of all KBOs is kept
at the usual value of 0.04, except for the binaries.
This is quite difficult to justify.
If, as seems more reasonable, all albedos have to be increased by a factor
of 4, then the size distribution of KBOs would keep its shape, but shifted
to sizes half the usual ones.
Finally, the number of potential impactors with the required mass would be
essentially the same as with the classical computation.
In such a scenario, we only gain because the critical specific
energy Q$^*_D$ is an increasing function of mass.

We have seen that if the much denser initial environment lasts long
enough to allow the elimination of most of the mass of the Kuiper
belt through collisional grinding (Davis and Farinella, 1997; Stern
and Colwell, 1997), then all the large separation KBO binaries would
be disrupted.
Simultaneously, many of the small separation binaries would also
be disrupted.
Since it seems impossible to create the large separation binaries in
the current environment, this implies that the dense initial
environment did not last long enough to allow for collisional
grinding.
Futhermore, collisional grinding can be effective at removing mass only for
very steep slopes ($b\la-4.5$) down to very small sizes, which is steeper
than that predicted by accretion models (Davis \al, 1999; Kenyon and Luu,
1999), at least in the 1--10~km range.
Hence, the mass reduction of the Kuiper belt by a factor of 100 must
result from a dynamical mechanism, as proposed by Gomes (2003) and
Levison and Morbidelli (2003).
This explanation is actually supported by the current inclination of
the KBOs which is much larger than their eccentricity on average (Gladman
\al, 2001).

This also allows to address the main criticism of the
Weidenschilling scenario by Golreich \al\ that Weidenschilling did
not propose a mechanism for disposing of the surplus of large bodies
needed at the beginning.
If it is the case that dynamics rather than collision erosion is
responsible for eliminating most the Kuiper belt mass, then the belt
erosion is independent of the size of the objects.
In such a case, one can easily dispose of 99\% of the belt mass, and
in particular of numerous large bodies.

\section{Summary}
\label{s:summary}

In this paper, we have shown that the stability of the KBO binaries with
respect to perturbations other than the usual shattering and dispersing of
fragments had to be investigated.
Ejection of the secondary from its orbit around the primary by a direct
collision of a rather small impactor turns out to be an efficient way to
eliminate KBO binaries.

The lifetime of 2001~QW$_{322}$ is 1 to 2~Gy, and hence there must have
been at least 10 times more similarly widely separated binaries formed at
the beginning.
Therefore, both Goldreich \al\ (2002) and Weidenschilling (2002) scenarios
of primordial formation must have been acting, Goldreich \al\ mechanism
forming most of the close binaries, Weidenschilling's forming most of the
large separation binaries.

Unless one can find a viable mechanism for forming numerous KBO binaries in
the current dynamical and collisional environment, we also showed that the
erosion of the Kuiper Belt cannot be due to collisional grinding.
It has to be the result of some dynamical effect that occured on a time
scale shorter than that necessary to collisionally erode most of the belt.

\section*{Acknowledgements}

We thank M. Nolan for discussions at ACM2002 that initiated this
work. We are also grateful to A. Morbidelli and B. Gladman who made
intereting suggestions that helped us improve our first manuscript.

\newpage

\begin{table}[h]
\begin{center}
\begin{tabular}{|l|c|c|c|c|c|c|c|}
\hline
\ & \multicolumn{3}{c|}{Merline \al \ (2003)} &
\multicolumn{3}{c|}{Stern (2002)} & \\
Object & R$_{P}$ & R$_{S}$ & Separ. & R$_{P}$ & R$_{S}$ & Separ. & $\Delta$ \\
name & (km) & (km) & (km) & (km) & (km) & (km) & (AU) \\[0.2cm]
\hline
1998 WW$_{31}$  &  75 &  60 &  22,300 & 169 & 141 &  22,000 &\ \  44.95\ \  \\
2001 QT$_{297}$ &\ \  290\ \  &\ \  207 \ \ &  20,000 &\ \  223\ \  & 173 &  20,000 & 44.80 \\
2001 QW$_{322}$ & 100 & 100 &\ \  130,000\ \  &  78 &  65 &\ \  130,000\ \  & 44.22 \\
1999 TC$_{36}$  & 370 & 132 &   8,000 & 293 & 107 &  11,000 & 39.53 \\
1998 SM$_{165}$ & 225 &  94 &   6,000 & 194 &  81 &   8,000 & 47.82 \\
1997 CQ$_{29}$  & 150 & 150 &   5,200 & 177 & 177 &   5,600 & 45.34 \\
2000 CF$_{105}$ &  85 &  53 &  23,000 & 117 &  79 &  23,000 & 44.20 \\
\hline
\ & \multicolumn{3}{c|}{$\Delta$ M = 0} &
\multicolumn{3}{c|}{$\Delta$ M = 2.2} & \\
\hline
2001 QC$_{298}$ &\ \  150\ \  &\ \  150 \ \ &  5,000 &\ \  199\ \  & 72 &  5,000 &  41.04 \\
\hline
\end{tabular}
\caption{
Characteristics of the binaries, according to Merline \al\
(2003) (columns 2 to 4), and Stern (2002) (columns 5 to 7). R$_{P}$ is the
radius of the primary, R$_{S}$ the radius of the secondary, $\Delta$ the
heliocentric distance, and columns 4 and 7 give the distance between the 2
components of the binary.
Values for 2001 QC$_{298}$ are given assuming equal size of both components
(columns 2 to 4) or a difference in magnitude of 2.2 (columns 5 to 7).
The assumed albedo is 0.04 in all cases, except for 1998~WW$_{31}$, for
which Merline \al\ assumed an albedo of 0.054.
Stern used a density of 2~g/cm$^{3}$, while Merline \al\ assumed a
more conventional density of 1~g/cm$^{3}$.
}
\label{t:characteristics}
\end{center}
\end{table}

\newpage

\begin{table}[ht]
\begin{center}
\begin{tabular}{|l|c|c|c|c|c|c|c|c|c|c|c|c|}
\hline
\ & \multicolumn{2}{c|}{Shattering} & \multicolumn{2}{c|}{Collisional} &
\multicolumn{2}{c|}{Gravity} & \multicolumn{2}{c|}{Shattering} & \multicolumn{2}{c|}{Collisional} &
\multicolumn{2}{c|}{Gravity} \\
\ & \ & \ & \multicolumn{2}{c|}{unbinding} & \ & \ & \ & \ & \multicolumn{2}{c|}{unbinding} & \ & \ \\
Object & R$_{shat}$ & T$_{shat}$ & R$_{h}$ & T$_{h}$ & R$_{g}$ & T$_{g}$ & R$_{shat}$ & T$_{shat}$ & R$_{h}$ & T$_{h}$ & R$_{g}$ & T$_{g}$ \\
name & (km) & Gyr & (km) & Gyr & (km) & Gyr & (km) & Gyr & (km) & Gyr & (km) & Gyr \\[0.2cm]
\hline
WW$_{31}$  &  18.5 &   38 &  6.3 &  1.3 & 288 &   20   &  80.5 &  830 & 25   & 25    & 492 &   59   \\
QT$_{297}$ & 108.5 & 1200 & 43   & 72   & 620 &  250   & 109.5 & 1500 & 35.5 & 54    & 492 &  100   \\
QW$_{322}$ &  38   &  150 &  9.5 &  1.9 & 620 &    3.4 &  26   &   96 &  6   &  0.85 & 492 &    1.3 \\
TC$_{36}$  &  56.5 &  330 & 34.5 & 75   & 620 & 1200   &  54   &  380 & 26.5 & 46    & 492 &  320   \\
SM$_{165}$ &  34.5 &  130 & 20   & 24   & 288 &  850   &  36   &  180 & 17   & 19    & 492 &  270   \\
CQ$_{29}$  &  68   &  470 & 30   & 39   & 288 &  630   & 113.5 & 1600 & 42.5 & 89    & 492 &  570   \\
CF$_{105}$ &  15.5 &   27 &  5.7 &  1.1 & 288 &   21   &  34.5 &  160 & 11.5 &  5.2  & 492 &   29   \\
\hline
\end{tabular}
\caption{
Minimum size of impactor (even columns) and corresponding
lifetime (odd columns) for the KBO binaries, for each of the three
disruption mechanisms: shattering, hitting, gravity perturbation.
Binary parameters correspond to Merline \al\ (2003) for columns 2 to 7, and
to Stern (2002) for columns 8 to 13.
}
\label{t:lifetimes}
\end{center}
\end{table}

\newpage

\begin{table}[htbp]
\begin{center}
\begin{tabular}{|l|c|c|c|c|c|c|c|c|}
\hline
\ & \multicolumn{4}{c|}{$\rho$  = 1 g cm$^{-3}$} &
\multicolumn{4}{c|}{$\rho$  = 2 g cm$^{-3}$} \\[0.2cm]
\ & \multicolumn{2}{c|}{$\Delta$ M = 0} &
\multicolumn{2}{c|}{$\Delta$ M = 2.2} & \multicolumn{2}{c|}{$\Delta$ M = 0} &
\multicolumn{2}{c|}{$\Delta$ M = 2.2} \\
Disruption & R$_{shat}$ & T$_{shat}$ & R$_{h}$ & T$_{h}$ & R$_{g}$ & T$_{g}$ & R$_{g}$ & T$_{g}$ \\
mechanism & (km) & Gyr & (km) & Gyr & (km) & Gyr & (km) & Gyr \\[0.2cm]
\hline
Shattering &  68 & 470 &  24 &  63 &  88.5 & 990 &  30.5 & 130 \\
Hitting    &  30 &  40 &  15 &  15 &  33.5 &  57 &  17   &  21 \\
Gravity    & 288 & 650 & 288 & 790 & 492   & 520 & 492   & 580 \\
\hline
\end{tabular}
\caption{
Same as Table~\ref{t:lifetimes}, but for the various parameters for
2001~QC$_{298}$.
}
\label{t:lifetimes-qc298}
\end{center}
\end{table}

\newpage

\newpage

\newpage

\end{document}